\documentclass[reprint,aip, superscriptaddress, amsmath,amssymb, prl, floatfix]{revtex4-2}
\usepackage{graphicx}
\usepackage{dcolumn}
\usepackage{bm}
\usepackage{xcolor}
\usepackage{hyperref}
\usepackage{ulem}
\usepackage[mathlines]{lineno}
\usepackage{tikz,xcolor,hyperref}
\usepackage{orcidlink}
\usepackage{ulem}
\definecolor{lime}{HTML}{A6CE39}

\DeclareRobustCommand{\orcidicon}{
	\begin{tikzpicture}
	\draw[lime, fill=lime] (0,0) 
	circle [radius=0.16] 
	node[white] {{\fontfamily{qag}\selectfont \tiny ID}};
	\draw[white, fill=white] (-0.0625,0.095) 
	circle [radius=0.007];
	\end{tikzpicture}
	\hspace{-2mm}
}

\foreach \x in {A, ..., Z}{\expandafter\xdef\csname orcid\x\endcsname{\noexpand\href{https://orcid.org/\csname orcidauthor\x\endcsname}
			{\noexpand\orcidicon}}
}
\begin{document}

\title{Spectroscopy of the spin waves of a synthetic antiferromagnet grown on a piezoelectric substrate}
\author{G. Y. Thiancourt\orcidG{}}
\thanks{}

\author{S. M. Ngom}
\author{N. Bardou}
\affiliation{%
Université Paris-Saclay, CNRS, Centre de Nanosciences et de Nanotechnologies, 91120, Palaiseau, France
}%
\author{T. Devolder \orcidT{}}
\email{thibaut.devolder@cnrs.fr}
\affiliation{%
Université Paris-Saclay, CNRS, Centre de Nanosciences et de Nanotechnologies, 91120, Palaiseau, France
}%

\date{\today}

\begin{abstract}

Efficient coupling between magnons and phonons requires material platforms that contain magnetic multilayers with versatile high-frequency properties grown on piezoelectric substrates with large electromechanical coupling coefficients. One of these systems is the CoFeB/Ru/CoFeB Synthetic antiferromagnet grown on Lithium Niobate substrate. We investigate its microwave magnetic properties using a combination of ferromagnetic resonance and propagating spin wave spectroscopy, from which we extract the dispersion relation of the acoustic branch of spin waves. The frequency and the linewidth of this spin wave resonance, its field dependence and its dispersion relation indicate that the magnetic properties are as good as when grown on standard non-piezoelectric substrates, as well as being in line with theory. This new material platform opens opportunities to extend microwave acousto-magnonics beyond the use of single layer magnets.

\end{abstract}
\maketitle
\section{Introduction}
Devices exploiting surface acoustic waves (SAWs) play a significant role in numerous modern technological applications, ranging from wireless communications \cite{campbell_surface_1998} to medicine \cite{lange_surface_2008}, sensors \cite{pohl_review_2000}, and RF signal processing in telecommunications. 
However, they have certain limitations, such as a reciprocal propagation and the near-absence of tunability. 
Circumventing these limitations may be possible by coupling SAWs with magnons \cite{kittel_interaction_1958, dreher_surface_2012,kus_giant_2023,kus_nonreciprocal_2021,kus_symmetry_2021,huang_large_2024,xu_nonreciprocal_2020,yamamoto_interaction_2022} since the latter feature tunability and non-reciprocity. In particular, the spin waves (SWs) within magnetic bilayers have attracted  interest \cite{kwon_giant_2016} because of their convenient high frequency SWs with long propagation lengths. 
Ensuring the high quality of both the magnetic thin film and the piezoelectric substrate is crucial to benefit from the SAW-SW coupling. 

In this study, we investigate SWs within a synthetic antiferromagnet (SAF) grown on a piezoelectric lithium niobate substrate LiNbO$_3$. The SAF consists of two magnetic layers of CoFeB, separated by a non-magnetic spacer layer mediating antiferromagnetic coupling. 
The SAF exhibits two eigenmodes: the optical mode and the acoustic mode, characterized by out-of-phase and in-phase precession of the magnetizations, respectively \cite{stamps_spin_1994,nortemann_microscopic_1993, ishibashi_switchable_2020, gallardo_reconfigurable_2019,millo_unidirectionality_2023,thiancourt_unidirectional_2024}. It is known that the SW properties are very sensitive to the material structure, the roughness and the interdiffusion inside the stack \cite{mouhoub_exchange_2023}. Here we measure the properties of acoustic SWs using inductive techniques and demonstrate that it is possible to grow high-quality SAFs on high performing LiNbO$_3$ piezoelectric substrates.


\section{\label{sec:Propagating Spin Wave Spectroscopy (propre)}Materials and devices}
\begin{figure}
    \centering
    \includegraphics[width=1\linewidth]{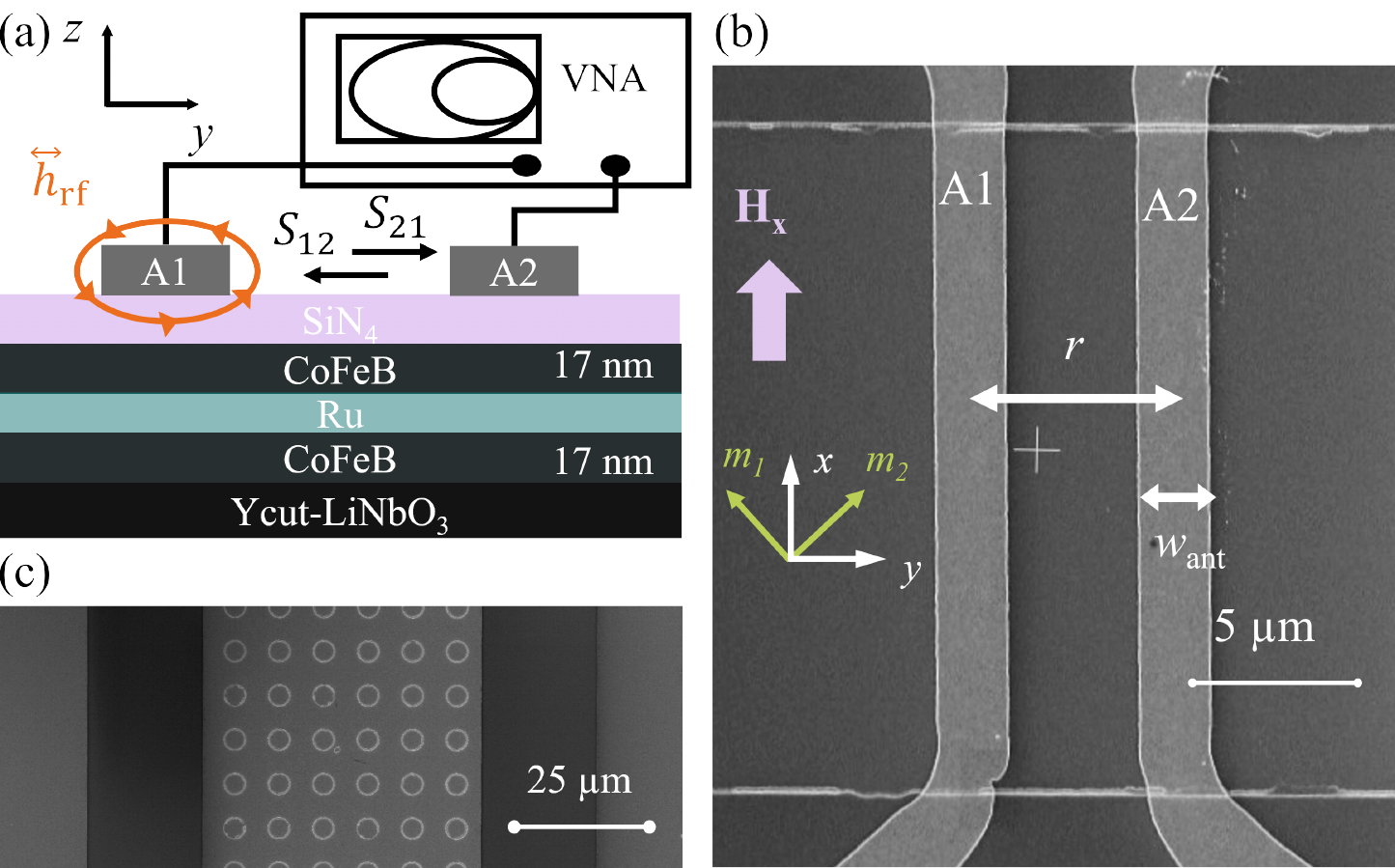}
    \caption{(a) Sketch of the setup of propagating spin wave spectroscopy (PSWS). (b) Scanning electron microscopy image of of a PSWS device. Light gray: single-wire antennas. Medium gray: Synthetic antiferromagnet conduit. (c) of array of SAF dots covered by a coplanar waveguide meant to measure VNA-FMR. }
    \label{fig:geometry}
\end{figure}

We use a SAF of composition LiNbO$_3$ (Y-cut substrate)/Ta /Co$_{40}$Fe$_{40}$B$_{20}$($t_\textrm{mag}=17$ nm) /Ru(0.7nm)
/Co$_{40}$Fe$_{40}$B$_{20}$($t_\textrm{mag}$) /Ru /Ta(cap) [Fig.~\ref{fig:geometry}.(a)]. The CoFeB has a magnetization $\mu_0M_s = 1.7~\textrm T$ and an interlayer exchange field $\mu_0H_j \approx 100$~mT. 
We pattern \cite{thiancourt_unidirectional_2024} the SAF into stripes of width $w_\textrm{mag}=20~\mu$m and much longer length. We cover the sample with $\textrm{Si}_3\textrm{N}_4$ and fabricate two micrometer-sized single wire antennas of widths $w_\textrm{ant}$ and variable center-to-center distances $r$ from 3.4 to 6~$\mu$m [Fig.~\ref{fig:geometry}(c)]. This device will be used to perform propagating spin waves spectroscopy (PSWS) and compare the material quality to existing benchmarks. The analysis of PSWS data will require the knowledge of the resonant frequencies $\omega(k=0)$ of uniform SWs, for which we fabricate an additional device to perform VNA-FMR \cite{bilzer_vector_2007}. To ensure comparability, the VNA-FMR device is built on the same chip as the PSWS device. It consists of an array of circular 4 $\mu $m dots \footnote{The geometry were chosen to reduce the effect of shape anisotropy.} inductively coupled to a much wider coplanar waveguide [Fig.~\ref{fig:geometry}(b)], which excites the uniform SW modes. The linewidth and the resonant frequencies of the uniform acoustic and optical spin wave modes shall provide the value of $H_j$ and of the damping parameter $\alpha$.



\section{Ferromagnetic resonance of the acoustic spin waves}



\begin{figure}
    \centering
    \includegraphics[width=1 \linewidth]{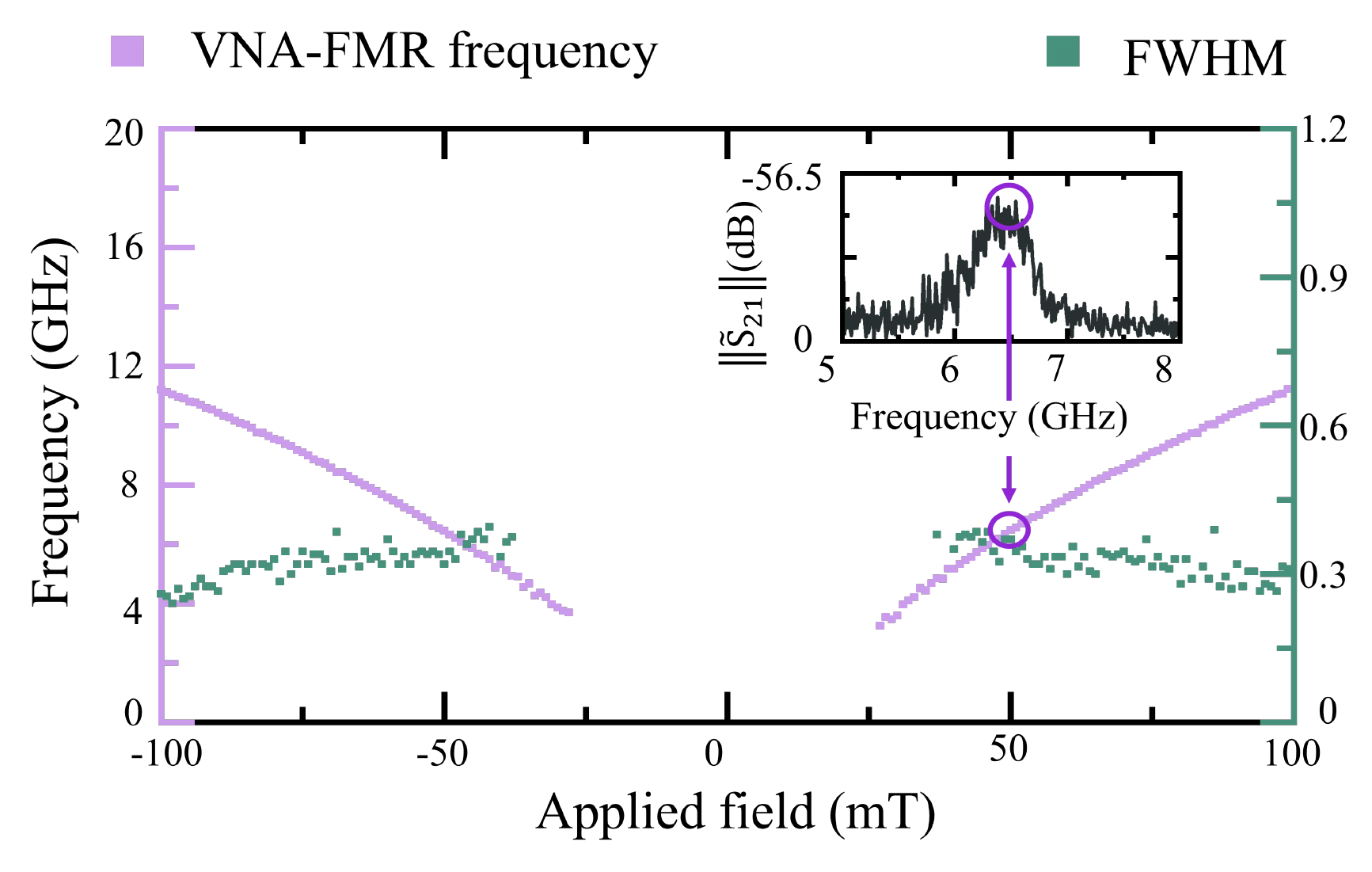}
    \caption{VNA-FMR measurements of the uniform acoustic spin wave resonance in the array of 4 $\mu$m dots. Left scale: Resonant frequencies as function of the applied field. The data correspond to the frequencies at which $\lvert \lvert \tilde S_{21} \rvert\rvert$ is maximal (inset). Right scale: Full widths at half maximum of these resonances.}
    \label{fig:FMR-frequency}
\end{figure}

Fig.~\ref{fig:FMR-frequency} summarizes the results obtained by VNA-FMR. The resonant frequency of the acoustic spin wave (pink curve) of the dots evolves quasi-linearly with the applied field. The resonance has the classical Lorentzian line shape (inset in Fig.~\ref{fig:FMR-frequency}), with a full width at half maximum that is almost independent of the field, being for instance 370~MHz for an applied field of $\mu_0H_x=50$~mT leading to a resonant frequency of $f_\textrm{dots}(k=0)=6.49$~GHz. The analysis of the linewidth using ref.~\onlinecite{devolder_measuring_2022} yields a Gilbert damping of $\alpha~=~ \Delta \omega_0 / (\gamma_0 (M_s + H_j))= 0.006 \pm 0.001$, where $\gamma_0$ is the gyromagnetic ratio.

Because of shape anisotropy, the uniform resonance of the dots differs from the uniform resonance of the stripe on which PSWS will be performed. Let us model this difference using the demagnetizing factors of our geometries $ \bar{\bar{N} }=\{N_{x},N_{y},N_{z}\}$. Using Eq.~11 of \onlinecite{beleggia_equivalent_2006}, the demagnetizing factors of each layer of the dot are $\bar{\bar{N} }_\textrm{dot}=\{0.01,0.01, 0.98\}$. Using \onlinecite{aharoni_demagnetizing_1998}, the demagnetizing tensor of each layer of the stripe is $\bar{\bar{N} }_\textrm{stripe}=\{0.004,0.001,0.995\}$. Since $N_{y}\leq N_{x}\ll N_{z}$, $\pm y$ is the easy axis and we can define a shape anisotropy field as $H_k=(N_x-N_y)M_s >0$. We also define the saturation fields $H_\textrm{sat,x}=H_j+H_k$ and $H_\textrm{sat,y}=H_j-H_k$ for saturation by fields in either the $x$ or the $y$ direction. 

For a field of magnitude  $H_k < H_x < H_\textrm{sat}$ applied in the $x$ direction, the SAF stripe is in a scissors state with the layers' magnetizations being:
\begin{equation} \left\{\begin{split}
    & m_{x1}=m_{x2}=\frac{H_x}{H_\textrm{sat,x}} \\
    & m_{y1}=-m_{y2} \\
    & m_{z1}=m_{z2}=0
\end{split} \right. \label{groudState} \end{equation} 
Linearization of the equations of motions can be used to derive the frequency of the acoustic spin wave:
\begin{equation} \begin{split}
   {\omega_\textrm{acou}}=   \gamma_0H_x \sqrt{H_j+(N_z-N_y)M_s} \\
   \times \sqrt{\frac{ H_\textrm{sat,x}}{H_\textrm{sat,y}^2}-\frac{H_k}{H_x^2}} \\
\end{split} \label{omegaAcou} \end{equation} 
 
We have deduced $f_\textrm{acou}^\textrm{stripe}(H_x)$ from $f_\textrm{acou}^\textrm{dots}(H_x)$, $\bar{\bar{N} }_\textrm{stripe}$, $\bar{\bar{N} }_\textrm{dot}$ and Eq.~\ref{omegaAcou}. The expected uniform resonance frequencies of the magnetic stripes for an applied field $\mu_0H_x=50$~mT is for instance $f_\textrm{acou}^\textrm{stripe}(H_x)=6.44$~GHz.
The resonance frequencies and their linewidth are very similar to that of samples grown on non-piezoelectric substrates \cite{seeger_inducing_2023, wojewoda_unidirectional_2023}. This is a first indicating that high quality SAF can be grown also on piezoelectric substrates. Let us determine the spin wave dispersion relation and its field dependence to assess this statement in a more quantitative way.

\section{Spin wave spectroscopy of the acoustic spin waves of the synthetic antiferromagnets}
\subsection{Propagating spin wave measurements and methods}
To get the dispersion relations, we performed propagating spin wave spectroscopy. We use the standard method and connect the two antennas A1 and A2 to the two ports of a calibrated VNA to measure the forward $\tilde S_{21}$ and backward $\tilde S_{12}$ transmission parameters for various applied field [Fig.~\ref{fig:geometry}(a)]. The SWs of SAF in this field configuration are known (and were checked) to be frequency-reciprocal \cite{millo_unidirectionality_2023}. The data processing meant to maximize the signal-to-noise ratio is carried out following the time-gating and field-differentiation methods detailed in refs.~\onlinecite{devolder_measuring_2021} and \onlinecite{thiancourt_unidirectional_2024}.

\begin{figure}
    \centering
    \includegraphics[width=1\linewidth]{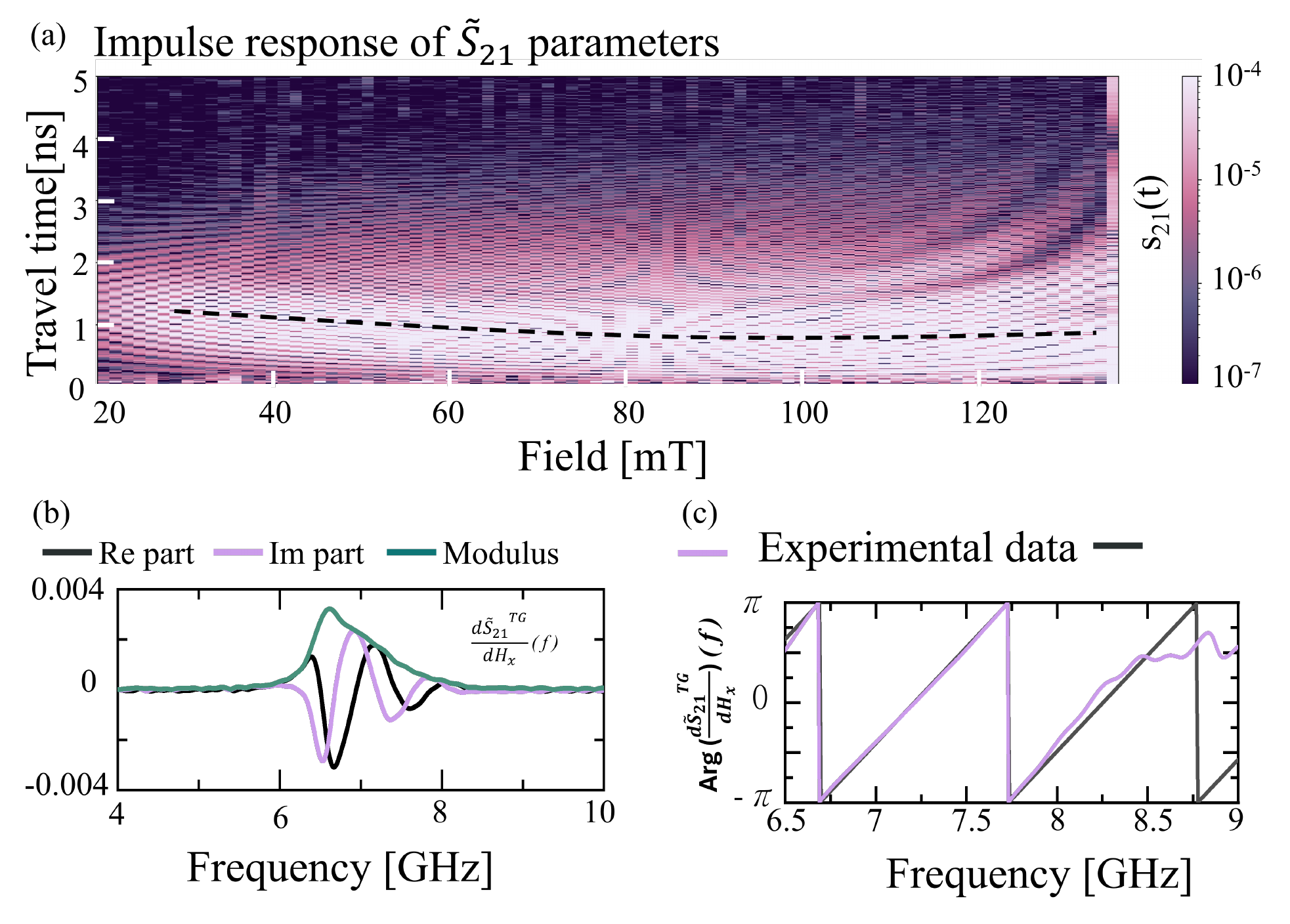}
    \caption{Propagating Spin Wave Spectroscopy of the acoustic spin waves channeled in the SAF stripe. (a) Impulse response of the antenna-to-antenna forward transmission coefficient. The black line is a guide to the eye that tracks the time at which the signal of the acoustic SW wavepacket is maximum. (b) Field derivative of the forward transmission parameter $\tilde S_{21}(f)$ for an applied field $\mu_0H_x =50$ mT after time-gating. (c) Comparison of the experimental phase at $\mu_0H_x =50$ mT with Eq. \ref{eq:Phase}. Starting from the experimental estimates of Eq.~\ref{omegaAcou}, the parameters were fitted to be $L_\textrm{att}= 3.6~\mu$m, $r=4.2~\mu$m, $v_g=4.4$ km.s$^{-1}$ and $f(k=0)=6.43$ GHz.}
    \label{fig:PSWS_measurement}
\end{figure}

Fig~\ref{fig:PSWS_measurement}.(a) shows the impulse response $s_{12}(t)$ for a SW propagation distance $r=~4.2~\mu$m. The wavepackets of the acoustic SWs reach the second antenna in a travel time of typically $t_{g} \approx 1$ ns. This group delay is essentially independent of the applied fields for fields inducing a scissor state \footnote{For applied fields near and above saturation $\mu_0H_x \geq 100$~mT, the initially single wavepacket of the impulse response progressively splits into several separated wavepackets arriving at different $t_g$'s. This is indicative to the presence of several SW mode with different group velocities, most probably quantized in the width of the stripe \cite{devolder_measuring_2022}. } [see the dashed line in Fig~\ref{fig:PSWS_measurement}(a)]. This group delay yields a first estimate of the group velocity of $v_g \approx r/t_g= 4.2~$km.s$^{-1}$ and a first estimate of the attenuation length of $L_\textrm{att}=\frac{2v_g}{\Delta\omega_0}=3.6~\mu$m of the propagating acoustic spin waves. 

Fig~\ref{fig:PSWS_measurement}(b) shows a representative processed transmission spectrum around the frequency of the acoustic spin waves. The phase of the signal is reported in Fig~\ref{fig:PSWS_measurement}(c). Transforming these phases into dispersion relations requires the modeling of the device's response.

\subsection{Model of $\tilde S_{21}$ for reciprocal spin waves}
For reciprocal waves with a $\vee$-shaped dispersion relation, the contribution $\tilde S_{21}^{yy}$ of the diagonal terms of the magnetic susceptibility to the forward transmission coefficient can be deduced from the sum of the experimental forward and reverse transmission coefficients (see the section IX.B of ref. \onlinecite{devolder_propagating-spin-wave_2023}):
\begin{equation} 
\tilde S_{21}^{yy}(\omega, r, \vee)=\frac 12 \Big(S_{21}^\textrm{exp}(\omega)+S_{12}^\textrm{exp}(\omega)\Big) 
\end{equation}
We consider in addition the limit of an ultra-thin antenna placed close to the magnetic film. Using the formalism of ref.~\onlinecite{devolder_propagating-spin-wave_2023}, we can reformulate an expression of $\tilde S_{21}(\omega)$ for reciprocal SWs with a $\vee$-shaped dispersion relation. 
Apart very near the uniform acoustic resonance, the scattering parameters can be expressed as \cite{devolder_propagating-spin-wave_2023}:
\begin{equation} \begin{split}
      \tilde S_{21}^{yy}(k > \frac{\Delta\omega_0}{v_g}, r, \vee) = \sqrt{\frac{\pi}{2}}\mathcal{L} (k)\Big(P_2(r-w_\textrm{ant},k) \\+P_2(r+w_\textrm{ant},k)-2 P_2(r,k)\Big)
\end{split} \label{S21yy} \end{equation}
for $k > \frac{\Delta\omega_0}{v_g}$, with $\mathcal{L}(k) \propto{ v_g^{-2} \left(k -i/ {L_\textrm{att}} \right)^{-2}}$ a Lorentzian function centered at $k=0$ and of width related to the inverse attenuation length \footnote{We neglect any variation of the attenuation length with the wavevector.}.
 The propagation term 
 is expressed using the function $P_2(x, k) =|L_{\textrm{att}}| e^{i\,k\, x } e^{-|\frac{x}{L_\textrm{att}}|} $ that expresses the phase rotation and the decay of the wave upon its propagation.
\\

Using Eq.~\ref{omegaAcou} and \ref{S21yy} and taking into consideration that $w_\textrm{ant}< r \ll L_\textrm{att}$, we can state that for  $k>\frac{2}{L_\textrm{att}}$ the experimental data can be linked to the SW wavevector of the acoustic SW branch by:
\begin{equation}
    \textrm{arg}\left(\frac{\partial \tilde{S}_{21}^{yy}}{\partial H_x} \left(\omega\right)\right)= 
    kr-\tan ^{-1}\left(\frac{kL_\textrm{att}}{4 }\right)+2\,n\,\pi, \label{eq:Phase}
\end{equation}
with $n\in \mathbb{Z}$. The value of $n$ can be found from the frequency of the uniform mode ($k=0$) determined in section II.
The equation \ref{eq:Phase} expresses that the phase of the field derivative of $\tilde S_{21}^{yy}(\omega)$ parameter evolves linearly with the wavevector with a distance independent correction term. This correction term can be rewritten as $\tan ^{-1}\left(\frac{\omega -\omega_0}{2 \Delta\omega_0 }\right)$ to evidence that it tends towards $\frac{\pi}{2}$ when working at frequencies $\omega > \omega_0 + 2 \Delta\omega_0 $. 
Fig.~\ref{fig:PSWS_measurement}(c) shows the comparison of the experimental phase of the field derivative of $\tilde S_{21}(f)$ with Eq.~\ref{eq:Phase} for input values adjusted according to experimental measurements. The agreement is satisfactory and the dispersion relation can thus be deduced from a fit to Eq.~\ref{eq:Phase}.

\subsection{Dispersion relation of the acoustic spin waves}
\begin{figure}
    \centering
    \includegraphics[width=0.8\linewidth]{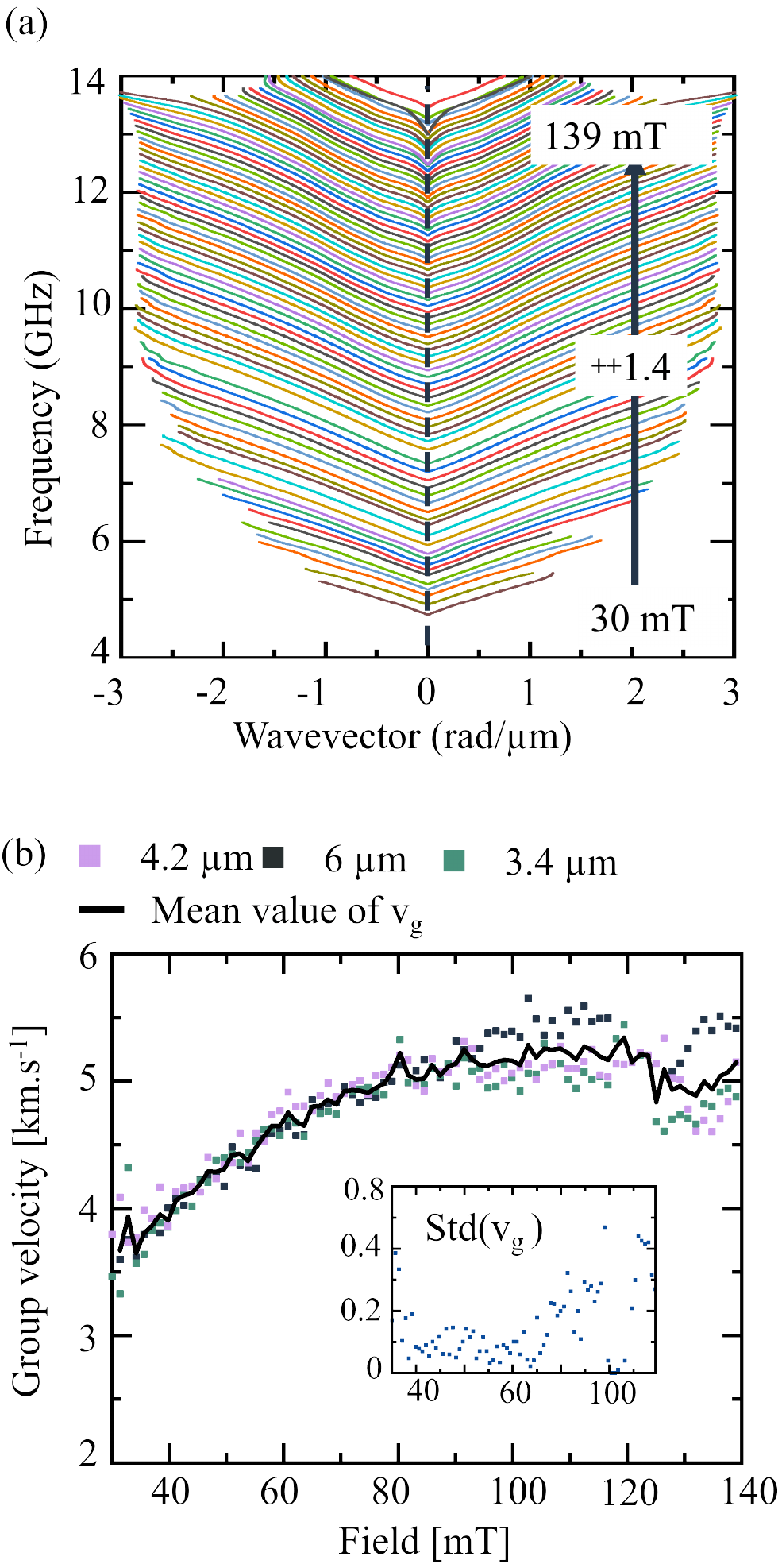}
    \caption{(a) Dispersion relation of the acoustic spin waves for $\vec{H}_x \perp \vec k$ obtained for a Propagating spin wave spectroscopy device with antenna width $w_\textrm{ant}=2~\mu$m and antenna to antenna distance $r=4.2~\mu$m. (b) Group velocities obtained for k=1 rad/$\mu$m and $r=4.2~\mu$m (violet), $r=6~\mu$m (black) and $r=3.4~\mu$m (green). Inset: standard deviation among these three measurements.}
    \label{fig:wk}
\end{figure}

Fig.~\ref{fig:wk}(a) gathers the dispersion relations of the acoustic SWs measured for applied field spanning from 30 mT up to 139 mT. This extends up to the maximum wavevector $k_\textrm{max} \approx 3~\textrm{rad}/\mu$m that can be investigated with our antenna width. The $k_y <0$  part of the dispersion relations are constructed by symmetry.

For each applied field, the quasi-linear V-shape of the dispersion relation allows to define a characteristic group velocity. Fig.~\ref{fig:wk}(b) reports the group velocity of SW deduced from linear fits done in the 0.7 to 1.3 rad/$\mu$m interval of wavevector. For $\mu_0H_x=50$~mT the group velocity is $4.422\pm0.062~\textrm{km.s}^{-1}$. It increases up to $5.220\pm0.041~\textrm{km.s}^{-1}$ for $\mu_0H_x=85$~mT. The group velocities deduced from samples with different propagation distances match together within a standard deviation of typically 200 m/s [inset of Fig.~\ref{fig:wk}(b)]. This consistency among different devices is indicative of a good homogeneity of the magnetic properties despite their growth on non-standard substrates. 

The group velocity is found to increase with the applied field until it saturates and decreases slightly for $\mu_0$H$_x > 100$~mT. This behavior is in semi-quantitative agreement with the predictions for a 2-macrospin SAF \cite{millo_unidirectionality_2023}, for which when $\vec{H}_x \perp \vec k_y$, the group velocity is expected to be $\frac{1}{2} \gamma_0 M_s t \frac{H_x}{H_j} \frac{M_s}{\sqrt{H_j(H_j+M_s)}}\,\textrm{sgn}(k_y) $ for $H_x < H_j$ (corresponding to the scissors state), and then to be slightly decreasing following $ \frac{1}{2} \gamma_0 M_s t  {\frac{M_s}{\sqrt{H_x(H_x+M_s)}}} \textrm{sgn}(k_y)$ for larger fields (corresponding to the saturated parallel state). This is an additional indication of proper synthetic antiferromagnetic behavior.



\section{Summary and concluding remarks}
In summary, we have measured several metrics representative of the dynamical properties of synthetic antiferromagnets grown on LiNbO$_3$ substrates and patterned into magnonic devices. The measurements include the uniform resonance of the acoustic spin waves, the dispersion relation of these spin waves, and consequently their group velocities and attenuation lengths. These metrics are indicative that the SAFs grown on LiNbO$_3$ have very similar properties as the SAFs grown on standard non-piezoelectric substrates, for instance on silicon substrates \cite{seeger_inducing_2023, millo_unidirectionality_2023}.
The new material platform shall enable to efficiently couple the magnetics degrees of freedom to the elastic ones, for instance to confer non-reciprocity to surface acoustic wave delay lines \cite{verba_wide_2019}.

\begin{acknowledgments}
This work was supported by the French RENATECH network, by the French National Research Agency (ANR) as part of the “Investissements d’Avenir” and France 2030 programs. This includes the SPICY project of the Labex NanoSaclay: ANR-10-LABX-0035, the MAXSAW project ANR-20-CE24-0025 and the PEPR SPIN projects ANR 22 EXSP 0008 and ANR 22 EXSP 0004. We acknowledge Aurélie Solignac for material growth and Loukas Kokkinos, Joo-Von Kim and Maryam Massouras for helpful comments.
\end{acknowledgments}

\section*{AUTHOR DECLARATIONS}
\subsection*{Conflict of Interest}
The authors have no conflicts to disclose.

\subsection*{Author Contributions}
\textbf{G.~Y.~Thiancourt}: Data curation (lead); Formal analysis (lead); Writing– original draft (equal); Writing– review and editing (equal); Measurements (supporting); Investigation (equal).
\textbf{S.~M.~Ngom}: Resources (equal); Investigation (equal); Measurements (equal).
\textbf{N.~Bardou}: Resources (equal).
\textbf{T.~Devolder}: Funding acquisition (lead); Investigation (equal); Measurements (equal); Writing– original draft (equal); Writing– review and editing (lead); Conceptualization (lead).

\section*{DATA AVAILABILITY}
The data that support the findings of this study are available within the article from the corresponding author upon reasonable request.

\section*{REFERENCES}
\bibliography{SAFPSWS}

\end{document}